\documentclass[useAMS,usenatbib]{mn2e} 
\usepackage{graphicx}

\usepackage[normalem]{ulem} 
\usepackage{color} 
\usepackage{amsmath} 
\usepackage{multirow}

\newcommand{\mt}{z_{p}\!\!:\!\!z_{d}}

\newcommand{\ximin}{\xi_{\rm min}} 
\newcommand{\mdot}{\langle \dot{M} \rangle} 
\newcommand{\mpc}{{\rm Mpc}} 
\newcommand{\kpc}{{\rm kpc}} 
\newcommand{\msun}{{\rm M_{\odot}}}

\title[Merger Rates]{The Merger Rates and Mass Assembly Histories of Dark Matter Haloes in the Two Millennium Simulations}

\author[O. Fakhouri, C.-P. Ma, M. Boylan-Kolchin]{Onsi Fakhouri$^{1}$\thanks{$\!\!$e-mail: onsi@berkeley.edu, cpma@berkeley.edu}, Chung-Pei Ma$^{1}$, and Michael Boylan-Kolchin$^2$\\
$^{1}$Department of Astronomy, 601 Campbell Hall, University of California, Berkeley, CA 94720\\
$^2$Max-Planck-Institut f\"{u}r Astrophysik, Karl-Schwarzschild-Str. 1, 85748 Garching, Germany} 
\begin{document}

\pagerange{ 
\pageref{firstpage} 
\pageref{lastpage}} \pubyear{2010}

\maketitle \label{firstpage} 
\begin{abstract}
  We construct merger trees of dark matter haloes and quantify their merger
  rates and mass growth rates using the joint dataset from the Millennium
  and Millennium-II simulations. The finer resolution of the Millennium-II
  Simulation has allowed us to extend our earlier analysis of halo merger
  statistics to an unprecedentedly wide range of descendant halo mass
  ($10^{10} \la M_0 \la 10^{15} M_\odot$), progenitor mass ratio ($10^{-5}
  \la \xi \le 1$), and redshift ($0 \le z \la 15$). We update our earlier
  fitting form for the mean merger rate {\it per halo} as a function of
  $M_0$, $\xi$, and $z$. The overall behavior of this quantity is
  unchanged: the rate per unit redshift is nearly independent of $z$ out to
  $z\sim 15$; the dependence on halo mass is weak ($\propto M_0^{0.13}$);
  and it is nearly a power law in the progenitor mass ratio ($\propto
  \xi^{-2}$). We also present a simple and accurate fitting formula for the
  mean mass growth rate of haloes as a function of mass and redshift. This
  mean rate is $46 M_\odot$ yr$^{-1}$ for $10^{12} M_\odot$ haloes at
  $z=0$, and it increases with mass as $\propto M^{1.1}$ and with redshift
  as $(1+z)^{2.5}$ (for $z\ga 1$). When the fit for the mean mass growth
  rate is integrated over a halo's history, we find excellent match to the
  mean mass assembly histories of the simulated haloes. By combining merger
  rates and mass assembly histories, we present results for the number of
  mergers over a halo's history and the statistics of the redshift of the
  last major merger.
\end{abstract}

\section{Introduction} \label{sec:intro} 

Mergers of dark matter haloes are intimately connected to a wide array of
phenomena in the now-standard $\Lambda$CDM cosmology. In addition to being
the dominant channel for mass growth of haloes themselves, mergers are also
responsible for the growth of stellar mass in galaxies, both directly via
galaxy-galaxy mergers, and indirectly via the accretion of potentially
star-forming gas. Furthermore, mergers help shape many important
observational properties of galaxies, e.g., star formation rates, color and
morphology transformations, dynamical states of stellar disks, and galaxy
mass and luminosity functions. Mergers are also responsible for the
existence of satellite galaxies such as dwarf spheroidals in the Milky Way
and non-cD galaxies in galaxy clusters. Quantifying the rate of halo-halo
mergers, and its possible dependence on factors such as halo mass, mass
ratio, and time, is therefore of great interest for a theoretical
understanding of galaxy formation and its connections to observations.

In a series of papers, we have examined various aspects of the growth of
dark matter haloes. In \citet{FM08}, we computed the merger rates of dark
matter haloes from the Millennium Simulation \citep{Springel05} and
presented a simple algebraic fitting form for our results. The resolution
and size of this simulation allowed us to determine the merger rate over
the parameter range of $10^{12} \la M_0 \la 10^{15} M_\odot$ for the mass
of the descendant haloes, $10^{-3} \la \xi \le 1$ for the mass ratio of the
progenitor haloes, and $0 \leq z \la 6$ for the redshift. The detailed
environmental dependence of the merger rates and halo mass growths was
analyzed in two subsequent papers \citep{FM09, FM10}. In \citet{MFM09}, we
studied the statistics of the halo mass assembly histories and mass growth
rates in the Millennium Simulation.  Halo mergers have also been
  studied in a handful of papers by others (e.g., \citealt{Governato99,
    Gottlober01, Berrier06, Maller06, Guo08, Genel09, Stewart09}).  The
  pre-2008 studies were all limited to small simulations that mainly
  investigated major mergers in a narrow mass range at low redshift
  (typically $z\la 1$).  Some such studies have emphasized potential
challenges for hierarchical structure formation; for instance,
\cite{Stewart08} have noted that the frequency of major mergers among
Milky-Way sized haloes poses a problem for thin-disk survivability.  Much
work has also been done in quantifying halo mass accretion and assembly
histories using $N$-body simulations that are smaller than the Millennium
runs (e.g., \citealt{LC94, Tormen97, Tormen98, WechslerMz, VDBosch, Li07,
  Zhao09}, except \citealt{Cole08}).

In this paper, we extend the results presented in \cite{FM08} and
\cite{MFM09} by incorporating the Millennium-II Simulation \citep{BK09}.
This simulation has the same number of particles as the Millennium
Simulation but has 125 times better mass resolution. This new database
provides $7.5\times 10^{6}$ dark matter haloes (each containing more than
1000 simulation particles) between redshift 0 and 15 and their subhalo
merger trees for our analysis. Adding to the $11.3\times 10^{6}$ haloes
(between $z=0$ and 6) available from the Millennium Simulation, this
combined dataset allows us to determine the dark matter halo merger rates
and mass growth rates from $z=0$ to up to $z=15$, for over five orders of
magnitude in the descendant halo mass ($10^{10} \la M_0 \la 10^{15}
M_\odot$) and progenitor mass ratio ($10^{-5} \la \xi \le 1$).

This paper is organized as follows. Section~\ref{sec:millennium} describes
the dark matter haloes in the Millennium and Millennium-II simulations, and
how we construct the merger trees and quantify the merger statistics and
mass accretion histories of the haloes. In Section~\ref{sec:results}, we
present results for three types of statistics: merger rates at $z=0$ up to
$\sim 15$ for halo mass $\sim 10^{10}$ to $10^{15} M_\odot$
(\S~\ref{sec:mergers}); the rate at which the haloes are accreting dark
matter across the virial radii, and the mass growth history of haloes
(\S~\ref{sec:MAH}); and the cumulative merger statistics over a halo's past
history, e.g., the mean cumulative number of mergers of a given mass ratio
experienced as a function of $z$ and halo mass, and the distribution of the
redshift at which the last major merger occurred for haloes at various
mass and redshift (\S~\ref{sec:mergerHistory}). The Appendix contains a
detailed comparison of the three types of algorithms that we have tested
for handling the fragmentation events in a merger tree of FOF haloes
\citep{FM08, FM10}. They are named ``snip,'' ``stitch,'' and ``split,''
depending on whether the fragmented subhalo was ignored, stitched back to
the original FOF halo in subsequent outputs, or split off from the FOF at
earlier times. A quantitative assessment of the systematic differences in
the merger rates derived from each algorithm is provided in the Appendix.

The cosmology used throughout this paper is identical to that used in the the Millennium simulations: a $\Lambda\textrm{CDM}$ model with $\Omega_m=0.25$, $\Omega_b=0.045$, $\Omega_\Lambda=0.75$, $h=0.73$, an initial power-law index $n=1$, and $\sigma_8=0.9$. Masses and lengths are quoted in units of $\msun$ and Mpc without the Hubble parameter $h$.

\section{Construction of Halo Merger Trees} \label{sec:millennium} 

\subsection{The Two Millennium Simulations} \label{sec:sub}

The Millennium and Millennium-II simulations are large $N$-body simulations of cosmological structure formation using the concordance $\Lambda$CDM cosmological parameters listed at the end of Section~\ref{sec:intro}. The simulations are described in detail in \cite{Springel05} (Millennium) and \cite{BK09} (Millennium-II); here we summarize some basic features of the simulations and of the default post-processing procedures that result in subhalo merger trees.

Both simulations follow the evolution of $2160^3 \approx 10^{10}$ particles from redshift 127 to redshift 0 using versions of the {\tt GADGET} tree-PM code \citep{Springel01, Springel05gadget}. The simulations differ in spatial scale and mass resolution: the Millennium Simulation uses a box size of $L=685 \,\mpc$ and a Plummer-equivalent force softening that is a factor of $10^{5}$ smaller, $\epsilon=6.85 \,\kpc$, with a particle mass of $m_p=1.18 \times 10^9 \,\msun$. The Millennium-II Simulation uses $L=137 \,\mpc$ and $\epsilon=1.37 \,\kpc$, both of which are a factor of 5 smaller than the values from the Millennium Simulation; the particle mass is therefore 125 times smaller, $m_p=9.43 \times 10^6 \,\msun$. The two simulations have 60 outputs at identical redshifts between $z \approx20$ and $z=0$, spaced approximately equally in $\log z$, as well as additional snapshots (4 for the Millennium, 8 for the Millennium-II) at higher redshifts.

Subhalo merger trees are constructed in an identical fashion for the Millennium and Millennium-II simulations. Dark matter haloes are first identified at each snapshot using a Friends-of-Friends group-finder (FOF; \citealt{Davis85}) with a linking length of 0.2 times the mean interparticle separation. All FOF groups with at least 20 particles are stored. The {\tt SUBFIND} algorithm \citep{springel01SUBFIND} is then applied to each FOF group to identify halo substructure. {\tt SUBFIND} identifies local density maxima and performs an unbinding procedure to determine which particles in the FOF group are bound to each density peak. Substructures with at least 20 particles after unbinding are stored, resulting in a list of subhaloes (SHs) associated with each FOF group in the simulation. Note that some FOF groups do not contain 20 self-bound particles and therefore not every FOF group contains a subhalo, while some FOF groups can contain many self-bound density peaks and therefore have many subhaloes.

These subhaloes are then linked across simulation snapshots to produce subhalo merger trees. This linking is done by establishing a unique descendant for each subhalo in the following manner. First, all particles in the subhalo are rank-ordered by binding energy. A list of candidate descendants -- all subhaloes at the subsequent snapshot containing at least one particle from the subhalo in question -- is built and a figure of merit is computed for each descendant. The candidate descendant with the highest score is assigned as the actual descendant. The figure of merit for candidate descendants is simply a weighted sum of the rank-ordering of the subhalo's particles; this procedure ensures that the tightly bound center of a subhalo is weighted more heavily than the less-bound outer regions even if the center is subdominant in terms of mass.

In addition to searching for a descendant at the subsequent output, a
search is also performed two snapshots later. This additional step accounts
for subhaloes that are temporarily unresolved when passing near the center
of a more massive system but re-appear later. On occasion, no descendant
can be identified at either of the two subsequent snapshots, in which case,
the subhalo is not assigned a descendant at all but rather is considered
destroyed. With subhaloes and their unique descendants identified, subhalo
merger trees are built by linking subhaloes and their descendants: all
subhaloes with a common descendant at $z=0$ are linked to all subhaloes
sharing these subhaloes as descendants, and so on. A given subhalo merger
tree thus contains all subhaloes that can be linked via their descendants
to one specific subhalo at $z=0$. The trees link 760 million subhaloes for
the Millennium Simulation and 590 million subhaloes for the Millennium-II
Simulation.  For the central subhalo of a $z=0$ galaxy-mass halo ($M
\approx 10^{12}\,\msun$), its subhalo merger tree typically consists of 90
subhaloes in the Millennium Simulation and 2800 subhaloes in the
Millennium-II Simulation.

\begin{table*}
	\centering 
	\begin{tabular}
		{l|l|cc|cc|cc|cc|cc|l} \multirow{2}{*}{$\mt$} & \multirow{2}{*}{Sim} & \multicolumn{2}{c}{$10^{10}\!-\!10^{11}M_\odot$} & \multicolumn{2}{c}{$10^{11}\!-\!10^{12} M_\odot$} & \multicolumn{2}{c}{$10^{12}\!-\!10^{13}M_\odot$} & \multicolumn{2}{c}{$10^{13}\!-\!10^{14}M_\odot$} & \multicolumn{2}{c}{$>\!10^{14}M_\odot$} & \multirow{2}{*}{Total}\\ &&$N_p=1$&$N_p\geq2$&$N_p=1$&$N_p\geq2$&$N_p=1$&$N_p\geq2$&$N_p=1$&$N_p\geq2$&$N_p=1$&$N_p\geq2$ \\
		\hline \multirow{2}{*}{0.12:0.06} & M & 0 & 0 & 0 & 0 & 321,489 & 90,922 & 14,504 & 45,281 & 3 & 4,817 & 477,016 \\
		& M-II & 214,045 & 25,292 & 12,583 & 17,107 & 13 & 3,279 & 0 & 486 & 0 & 36 & 272,841 \\
		\hline \multirow{2}{*}{0.56:0.51} & M & 0 & 0 & 0 & 0 & 306,142 & 98,664 & 11,442 & 39,469 & 0 & 2,757 & 458,474 \\
		& M-II & 224,865 & 29,170 & 12,422 & 18,405 & 7 & 3,199 & 0 & 421 & 0 & 20 & 288,509 \\
		\hline \multirow{2}{*}{1.17:1.08} & M & 0 & 0 & 0 & 0 & 236,280 & 137,729 & 4,197 & 32,349 & 0 & 976 & 411,531 \\
		& M-II & 220,703 & 49,811 & 8,473 & 23,221 & 1 & 2,985 & 0 & 316 & 0 & 8 & 305,518 \\
		\hline \multirow{2}{*}{2.23:2.07} & M & 0 & 0 & 0 & 0 & 126,926 & 133,965 & 629 & 12,746 & 0 & 73 & 274,339 \\
		& M-II & 202,572 & 80,435 & 4,772 & 24,874 & 2 & 2,128 & 0 & 121 & 0 & 0 & 314,904 \\
	\end{tabular}
	
        \caption{ The number of merger events in the two Millennium simulations at four representative redshifts ($z\approx 0$, 0.5, 1, and 2). At each $z$, we list the number of descendant FOF haloes that have a single progenitor halo ($N_p=1$, i.e., no mergers) and multiple progenitors ($N_p\ge 2$), for five descendant mass bins (left to right). The descendant mass here refers to the halo mass at the redshift listed rather than at the present day.  Only haloes containing more particles than our minimum cutoff (1000 for descendants; 40 for progenitors) are counted. The higher-resolution Millennium-II Simulation dominates the contribution to the merger statistics of $M_0 \la 10^{12} M_\odot$ haloes, while the larger-volume Millennium Simulation dominates the contribution to cluster-mass haloes. } \label{table:NumProgs} 
\end{table*}

\subsection{Halo Fragmentation} \label{FOFLims}

In this paper, as in our previous work \citep{FM08, MFM09}, our focus is on the merger and assembly histories of FOF haloes. To do this we must first construct merger trees of the FOF haloes from the underlying subhalo trees described in Section~\ref{sec:sub}. Such construction is nontrivial due to halo fragmentations: subhaloes of a progenitor FOF halo may have descendants that reside in more than one FOF halo. Sometimes this is due to a physical unbinding event in which a subhalo formerly bound to an FOF is ejected out of the FOF system. Sometimes the fragmentation is spurious -- a subhalo may oscillate in and out of the FOF group before finally settling in. Sometimes the FOF algorithm incorrectly groups subhaloes that are unbound but only happen to pass by one another and should not be associated as a single FOF group.

We presented detailed comparisons in \cite{FM08, FM10} of three types of algorithms -- snip, stitch, and split -- for handling these fragmentation events. In the Appendix we summarize these algorithms and quantify the systematic differences in the merger rates derived from each algorithm.

For the main results presented in \S~\ref{sec:results} below, we use the split-3 algorithm, in which the subhalo fragments that pop out of an FOF halo are either snipped or split depending on a simple criterion. The fragmented subhalo is snipped if it is observed to remain in the FOF halo for all 3 snapshots immediately preceding the fragmentation event; in this case the ancestral link between fragment and FOF is severed. If the fragmented subhalo is not in the FOF halo for all 3 preceding snapshots, it is interpreted as distinct and is split off from the FOF. The split-3 algorithm generally gives very similar results to the stitch-3 algorithm used in \cite{FM08}, e.g., the two methods produce merger rates that agree to within 10\% for the redshifts and mass ranges that we have statistics for. The only exception is in the major merger regime for low-mass haloes at low redshift ($z \la 1$), where split-3 is lower than stitch-3 by up to 30\% (see Fig.~\ref{fig:BNRatio} in Appendix). Overall, split-3 appears slightly more robust at handling spurious FOF linking events in this regime (also see \citealt{FM10, Genel09}). As discussed in the Appendix, however, the exact definition of what constitutes a merger may be situation-dependent, meaning that no single method is perfect in all cases.

\subsection{Extracting Merger Rates and Mass Accretion Histories} \label{sec:extract}

From the merger trees of FOF haloes obtained by applying a given
fragmentation algorithm, we extract a merger catalog. Each catalog provides
us with a list of descendant FOF haloes at redshift $z_d \geq 0$ with mass
$M_0$, and for each descendant halo, its set of $N_p$ FOF progenitors at
$z_p=z_d+\Delta z$, where $N_p$ can range from 1 (i.e. a single progenitor)
to a large number, depending on the halo mass and the value of $\Delta z$.
We label the rank-ordered progenitor mass with $M_{i}$,
$i\in(1,2,\dots,N_{p})$, and $M_1\geq M_2 \geq \dots M_{N_p}$. To ensure
that only numerically resolved haloes are included in our study, we impose
a minimum of 1000 particles for the descendant haloes and 40 particles for
the progenitor haloes. For the Millennium Simulation, this criterion
corresponds to a minimum halo mass of $1.2\times10^{12} \,\msun$ for the
descendant and $4.7\times10^{10} \,\msun$ for the progenitor. For
Millennium-II, the minimum masses are $9.4\times 10^9 \,\msun$ and
$3.8\times 10^8 \,\msun$ for the descendant and progenitor haloes.
We emphasize that the mass of 
a descendant halo refers to its mass at a given redshift $z_d$ 
and not its ultimate mass at $z=0$ (unless $z_d=0$).

We compute the merger rates at redshift $z$ as a function of descendant
mass $M_0$ and progenitor mass ratio $\xi=M_i/M_1$ (for $i>1$). We define
$B(M_0,\xi,z)$ to be the number of mergers per $\mpc^3, \, dM_0, \,d\xi,\,
{\rm and}\,\Delta z$ with mass $M_0\pm dM_0/2$ and mass ratio $\xi\pm
d\xi/2$. As discussed in \cite{FM08}, we find the mean merger rate {\it per
  halo}, $B(M_0,\xi,z)/n(M_0,z) \equiv dN_m/d\xi/dz$, where $n(M_0,z)$ is the number
density of haloes, to have a particularly simple dependence on the merger
parameters. This rate, when expressed in per redshift units, is a
dimensionless quantity that gives the mean number of mergers per halo per
unit $z$ per unit $\xi$. To avoid artificial boundary effects at $z=0$, we
use the two outputs at $z=0.12$ and 0.06 to compute the $z\sim 0$ merger
rate.

To compute the mass accretion histories and accretion rates of haloes, we
start with a given descendant FOF halo at some redshift and identify the
mass of its most massive progenitor at an earlier snapshot. This process is
iterated backwards in time to construct the main branch of the descendant's
merger tree. The mass trajectory along the main branch of a descendant
gives us its mass accretion history $M(z)$ (see. e.g., \citealt{LC93}),
from which we can compute the mass accretion rate $\dot{M}$ as a function
of $z$. Note that the progenitor halo on the main branch of a descendant
halo at a given snapshot need not be the most massive progenitor of that
descendant at that snapshot.

\begin{figure*}
  \centering
  \includegraphics{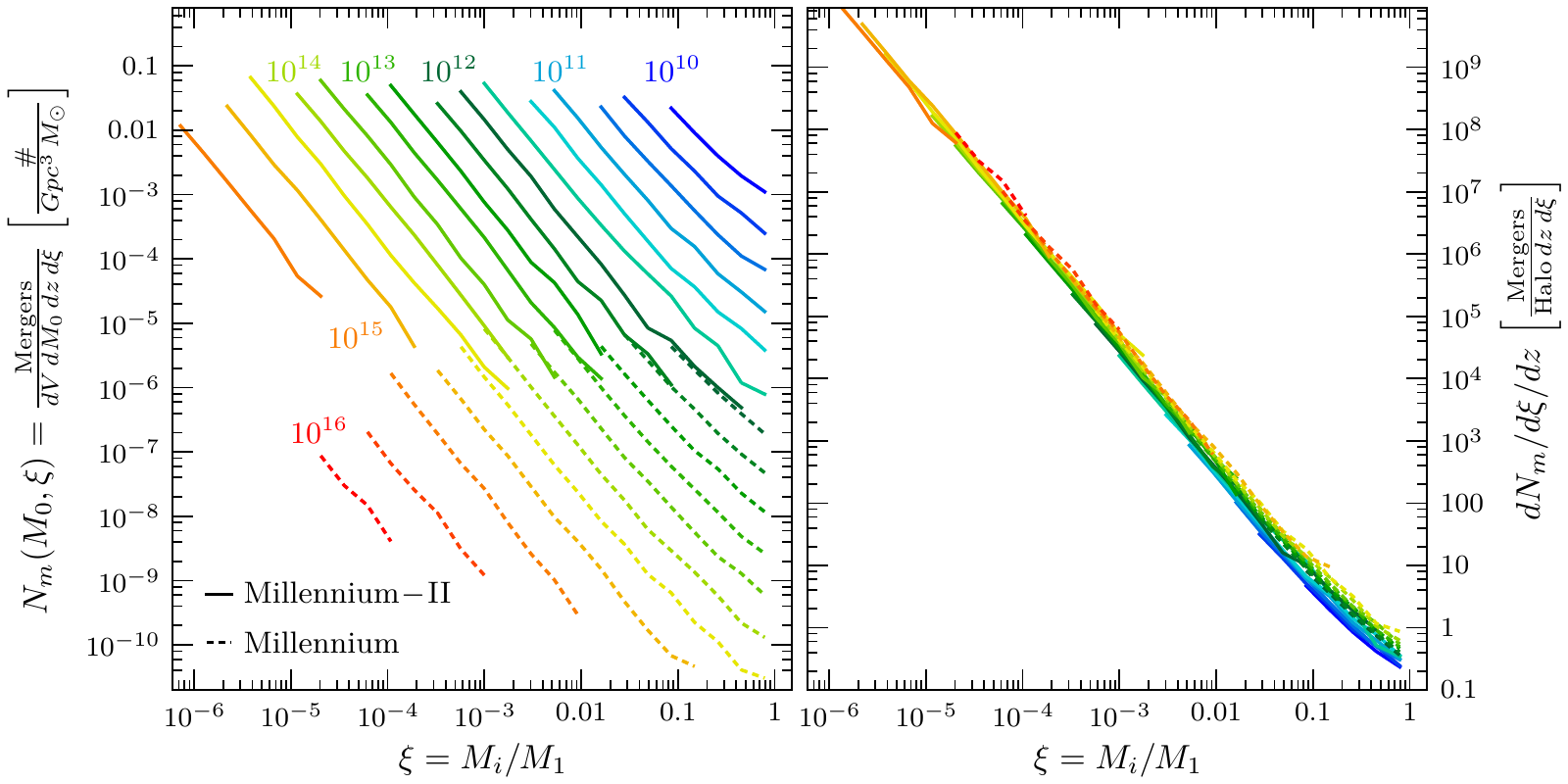}
  \caption{Left panel: The mean merger rate of $z=0$ FOF haloes,
    $B(M_0,\xi)$, as a function of the mass ratio of the progenitors
    ($\xi$) and the descendant halo mass ($M_0$) over 6 orders of
    magnitude: $10^{10}$ to $10^{16} M_\odot$ from right (blue) to left
    (red). The Millennium-II results are shown in solid, while the results
    from the Millennium are in dashed curves. Right panel: The mean merger
    rate {\it per halo}, $B(M_0,\xi)/n(M_0)\equiv
    dN_m/d\xi/dz$. Normalizing $B(M_0,\xi)$ by the halo number density
    $n(M_0)$ collapses the curves in the left panel to nearly a single
    curve, indicating that $dN_m/d\xi/dz$ is nearly independent of $M_0$
    and has a simple universal form.} \label{fig:Collapsing}
\end{figure*}

\subsection{Definitions of Halo Mass} \label{HaloMass}

In our prior analysis of the Millennium halo merger rate \citep{FM08}, we
assigned the halo mass using the standard FOF mass $M_{FOF}$. This mass is
simply proportional to the number of particles assigned to each FOF halo by
the FOF group finder. An alternative definition that we will use throughout
this paper is the sum of the masses of an FOF's subhaloes, $M_{SH}$. This
definition has been shown recently by \cite{Genel09} to be more robust than
$M_{FOF}$ since the {\tt SUBFIND} algorithm assigns only gravitationally
bound particles to each subhalo.

Overall, we find the halo mass functions computed using these two mass
definitions to differ at the 5\% level at all halo masses.  This
  difference can be caused by a slight excess of mass in FOF haloes due to
  unbound or spuriously linked particles, as well as by a slight deficit in
  $M_{SH}$ when {\tt SUBFIND} does not account for all the mass physically
  associated with a subhalo.  When restricted to the subset of haloes that
are undergoing very minor mergers, however, \cite{Genel09} noted that the
FOF mass of the smaller progenitor increases as it approaches the more
massive progenitor. For minor mergers involving mass ratios as low as
$\xi\sim 0.001$, the ratio $M_{FOF}/M_{SH}$ for the {\it smaller}
progenitor can rise from 1.03 to 1.5 prior to mergers. We will therefore
use $M_{SH}$ for halo masses in this study. We note that this discrepancy
occurs only for the small subset of low-mass haloes that are in the process
of merging onto a much larger halo; its effect on the total halo mass
function is therefore limited to $\sim 5$\%.

\section{Results} \label{sec:results}

To provide a sense for the halo statistics and merger events available from
the two Millennium simulations, we list in Table~\ref{table:NumProgs} the
number of descendant haloes (above 1000 particles) and their progenitors
(above 40 particles) at four representative redshifts for five broad mass
ranges from $10^{10}$ to $>10^{14}M_\odot$. The results presented below are
based on these events and those at other redshifts.

\subsection{Merger Rates} \label{sec:mergers}

\subsubsection{Present-Day Merger Rates} \label{sec:mergersToday}

The left panel of Fig.~\ref{fig:Collapsing} presents $B(M_0,\xi,z=0)$, the
$z=0$ mean number of mergers per unit volume, descedant mass $M_0$, mass
ratio$ \xi$, and redshift as a function of progenitor mass ratio $\xi$ from
the two Millennium simulations (solid for Millennium II; dashed for
Millennium). The colored curves correspond to different mass bins ranging
from $10^{10} \,\msun$ (blue) to $10^{16} \,\msun$ (red). The rates are
determined from the $z=0.06$ and 0.12 merger tree catalogue since, as
described in Section~\ref{sec:extract}, we would like to avoid the $z=0.0$
snapshot due to the boundary effects that interfere with the
post-processing algorithms used to handle the halo fragmentation
events. The split-3 algorithm is used here; other algorithms yield
qualitatively similar agreement between the two simulations (see Appendix
for details).

The right panel of Fig.~\ref{fig:Collapsing} shows the mean merger rate per
halo, $B/n=dN_m/d\xi/dz$, where each of the curves in the left panel has been
divided by the number density of haloes in that mass bin. The collapse of
the curves to nearly a single curve shows that the per halo merger rate
$dN_m/d\xi/dz$ is nearly independent of halo mass. This collapse is similar
to that seen in Fig.~6 of \cite{FM08} for the Millennium Simulation. A
comparison of the two figures helps illustrate the large dynamic range
achieved when the two Millennium simulations are combined: the halo mass
range has been increased by two orders of magnitude in
Fig.~\ref{fig:Collapsing}, and for each mass bin, the progenitor mass ratio
is extended downward by also a factor of $\sim 100$, reaching $\xi \sim
10^{-6}$ for $M_0=10^{15}\,\msun$.

The overlap in the merger parameter space between the two simulations is
seen to be fairly small in Fig.~\ref{fig:Collapsing}. The two simulations
are therefore quite complementary: Millennium II allows us to probe
descendant and progenitor masses that are a factor of 125 smaller than
Millennium, whereas the larger box of the Millennium Simulation provides
robust statistics for the rare events that are poorly sampled in Millennium
II, e.g., major mergers of massive haloes (i.e. the lower right corner of
left panel of Fig.~\ref{fig:Collapsing}). Over the small region of overlap,
Fig.~\ref{fig:Collapsing} shows good agreements between the merger rates
determined from the two simulations: both the power-law dependence on $\xi$
and the weak dependence on $M_0$ carry over from Millennium to
Millennium II. \cite{BK09} show that many other quantities, such as halo
mass functions, formation times, and subhalo abundances, have a much wider
range of overlap and that the two simulations are in excellent agreement
for these quantities as well.

\begin{figure}
	\centering 
	\includegraphics{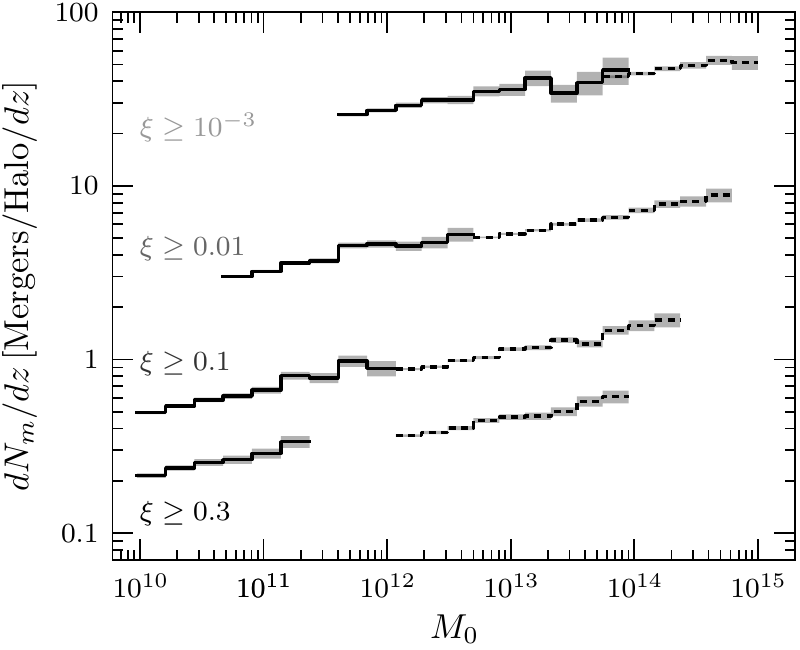} \caption{The $z\approx 0$ mean merger rate per halo (per unit $z$), $dN_{m}/dz$, as a function of descendant mass, $M_0$, for four ranges of progenitor mass ratio $\xi$. The Millennium-II results are shown in solid, while the original Millennium results are the dashed curves. The upper curves include increasingly more minor mergers. The mass dependence is weak over five orders of magnitude in mass and is well approximated by a power law $\propto M_0^{0.133}$. } \label{fig:MassDependence} 
\end{figure}

The weak dependence of the merger rate on $M_0$ is shown explicitly in
Fig.~\ref{fig:MassDependence}. Each curve here shows the mean rate per
halo, $dN_m/d\xi/dz$, integrated over different ranges of $\xi \ge \ximin$,
where $\ximin=0.3, 0.1 0.01$, and $10^{-3}$ (from bottom up). Major mergers
with mass ratio within 1:3 (bottom curve) are clearly much more rare than
minor mergers (top curves), but all the curves have very similar power-law
dependence on $M_0$. Over about 4.5 orders of magnitude in $M_0$, the rate
increases by only a factor of $\sim 3$, suggesting that the merger rate
scales roughly as $\sim M_0^{0.1}$. A more accurate fit is provided in
Sec~\ref{sec:fit} below.

\subsubsection{$z>0$ Merger Rates} \label{sec:mergersZ}

\begin{figure*}
	\centering 
	\includegraphics{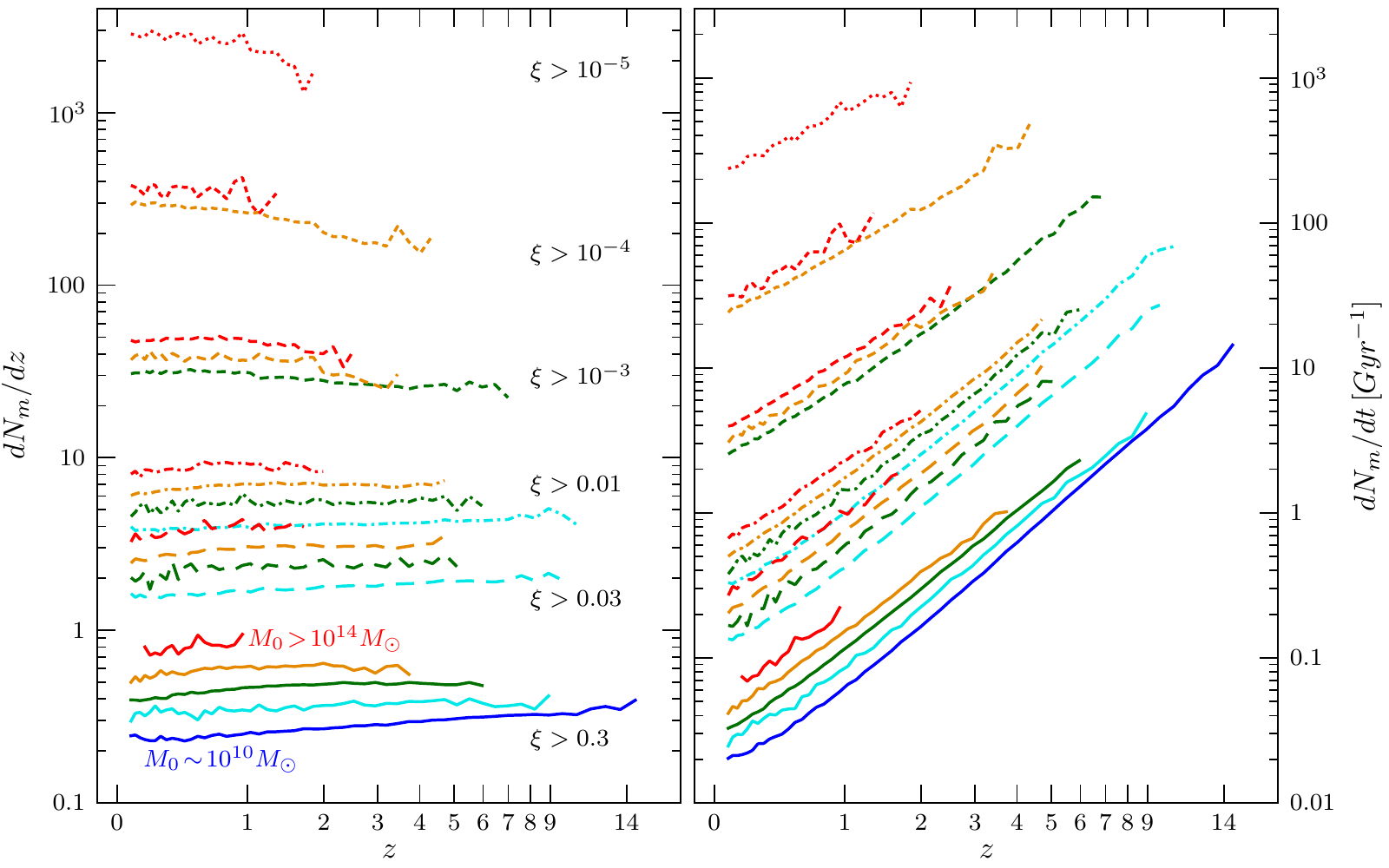} \caption{Time evolution of the
          mean halo merger rates {\it per halo} in units of per redshift,
          $dN_m/dz$ (left panel), and in units of per Gyr, $dN_m/dt$ (right
          panel) from the two Millennium simulations. The descendant mass
          $M_0$ and progenitor mass ratio $\xi$ over five orders of
          magnitude are plotted. The weak dependence of the rates on $M_0$
          is shown by the different colors: $\approx 10^{10}$ (blue),
          $10^{11}$ (cyan), $10^{12}$ (green), $10^{13}$ (orange). to $ >
          10^{14}\msun$ (red). The line types denote different types of
          mergers, ranging from major mergers (solid) to extreme minor
          mergers (dotted) The rate $dN_m/dz$ on the left is remarkably
          constant out to $z\sim 15$; the rapid rise of $dN_m/dt$ with
          increasing $z$ on the right is therefore largely due to the
          cosmological factor $dt/dz$, which spans a shorter time per unit
          $z$ with increasing $z$. } \label{fig:BNZ}
\end{figure*}

The Millennium Simulation provided sufficient halo statistics for us to
determine the halo merger rates up to $z\sim 6$ in our previous study. The
higher mass resolution of the Millennium-II Simulation now allows us to
probe redshifts up to $\sim 15$. The combined results from the two
simulations are shown in Fig.~\ref{fig:BNZ}, which plots the mean merger
rate per unit redshift (left panel), $dN_m/dz$, and per unit time (right
panel), $dN_m/dt$, as a function of redshift. These merger rates have been
integrated over different ranges of $\xi\ge \ximin$, ranging from major
mergers with $\ximin=0.3$ (solid curves at bottom), to extreme minor
mergers with $\ximin=10^{-5}$ (dotted curve at top). Within each line type,
the colors indicate different descendant mass bins ranging from $10^{10}$
(blue) to $> 10^{14}\msun$ (red). Only the higher mass bins are plotted as
$\ximin$ is lowered. This is because minor mergers of low-mass haloes fall
below the mass resolution limit.

Fig.~\ref{fig:BNZ} indicates that the general trends reported in Fig.~8 of
\cite{FM08} continue to hold in the Millennium-II Simulation. The
dimensionless rate $dN_m/dz$ is remarkably independent of redshift up to
$z\sim 15$, whereas the rate per Gyr, $dN_m/dt$, rises with increasing $z$
because a unit redshift corresponds to a shorter time interval at higher
$z$. This redshift dependence is similar to that obtained by \citet{Guo08}
for the dimensionless growth rates due to mergers of both haloes and
galaxies (based on semi-analytic models) from the Millennium Simulation.

\subsubsection{Merger Rate Fitting Forms} \label{sec:fit}

Since the merger statistics in Figs.~\ref{fig:Collapsing}-\ref{fig:BNZ} are
very consistent between the two Millennium simulations, we use an
analytical form similar to equation~(12) of \cite{FM08} to fit the
dimensionless mean merger rate $dN_m/d\xi/dz$ (in units of mergers per halo
per unit redshift per unit $\xi$) from the combined Millennium dataset. An
appealingly simple feature of this fitting form is that it is separable in
the variables $M_0,\, \xi$, and $z$:
\begin{equation}
	\frac{dN_m}{d\xi dz}(M,\xi,z)=A\left(\frac{M}{10^{12} M_\odot}\right)^\alpha
     \xi^\beta \exp\left[\left(\frac{\xi}{\tilde{\xi}}\right)^\gamma\right] (1+z)^\eta \,. 
\label{fit} 
\end{equation}
We find the best-fit parameters to be $(\alpha,\beta,\gamma,\eta)=(0.133,
-1.995, 0.263, 0.0993)$ and $(A,\tilde\xi)=(0.0104,
9.72\times10^{-3})$. The near $z$-independence in the left panel of
Fig.~\ref{fig:BNZ} is more striking than in our 2008 study due to the
larger coverage in redshift here. In view of this lack of $z$-dependence,
we choose to use the simpler factor of $(1+z)^\eta$ here rather than the
growth rate of the density field used in \cite{FM08}.  In comparison
  to our 2008 study, the power-law slope of the mass dependence is slightly
  steeper here ($\alpha=0.133$ vs. 0.089), whereas the power-law slope of
  the $\xi$ dependence is slightly shallower here ($\beta=-1.995$
  vs. $-2.17$).  These differences are primarily due to the differing
  definitions of halo mass used in the two studies (FOF vs. sum of
  subhalos; see Sec.~2.4) and the refinements in our stitch-3 algorithm
  (see the Appendix). 

We note that the left panel of Fig.~\ref{fig:BNZ} does show mild variations
in the redshift dependence among the different $\xi$ bins: the rate
increases slightly with increasing $z$ for major mergers, while it declines
somewhat for the very minor mergers ($\ximin \sim 10^{-4}$ to
$10^{-5}$). Since this variation is so minor and the minor merger regime is
more prone to numerical resolution issues, we have opted for simplicity
rather than a more complicated fitting form.

\subsection{Mass Growth Rates and Assembly Histories} \label{sec:MAH}

In the last section, we presented results for the instantaneous rates of
halo mergers as a function of redshift, descendant mass, and progenitor
mass ratio. Here, we examine a related set of statistics that quantify the
mass growth of haloes. These two quantities are clearly related since
mergers are a primary process for haloes to gain mass, but mergers are not
the only process. As discussed at length in \cite{FM10}, ``diffuse''
accretion of unresolved haloes or dark matter particles also makes an
important contribution to halo growth. In the mass assembly history of a
halo, mergers with other haloes typically result in more discrete but less
frequent changes in the halo mass, while diffuse accretion leads to a more
continuous change.

\subsubsection{Mass Accretion Rates}

\begin{figure}
  \centering
  \includegraphics{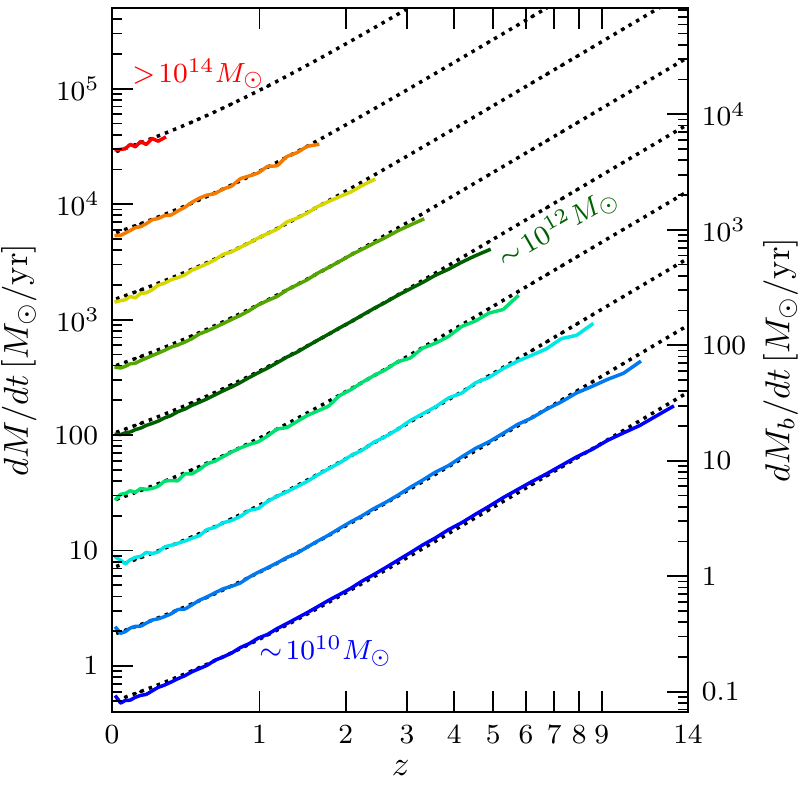} \caption{Mean mass accretion rate of
    dark matter onto haloes as a function of redshift from the two
    Millennium simulations (solid curves). Halo masses ranging from
    $10^{10} M_\odot$ to $>10^{14} M_\odot$ are plotted. The dashed curves
    show the accurate fit provided by equation~(\ref{Mdotfit}). The
    right-hand side of the vertical axis labels the mean accretion rate of
    baryons, $M_b$, assuming a cosmological baryon-to-dark matter ratio of
    1/6. } \label{fig:Mdot}
\end{figure}

\begin{figure*}
  \centering
  \includegraphics{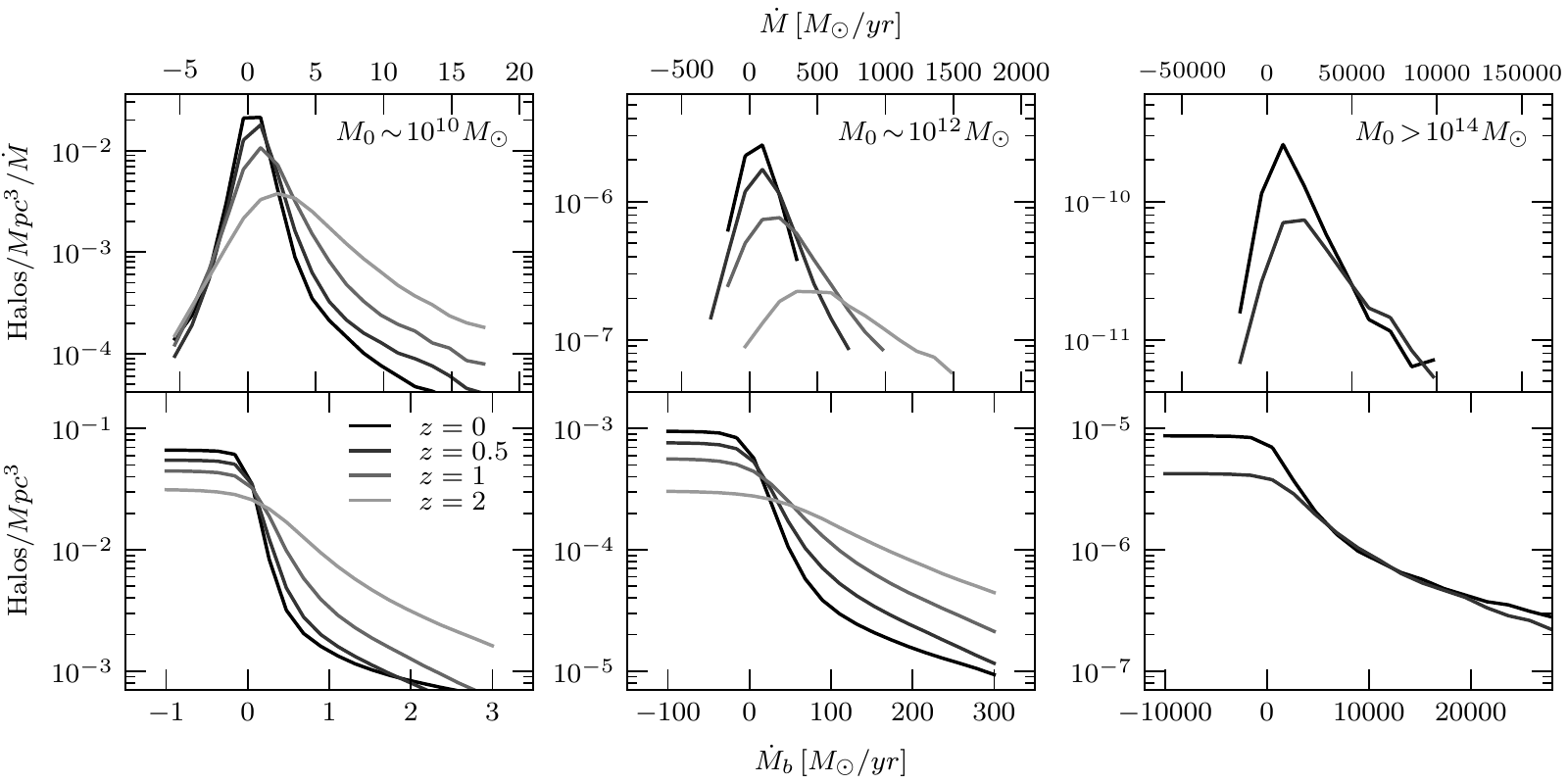} \caption{Differential (top) and
    cumulative (bottom) distributions of the baryonic accretion rates,
    $\dot{M}_b$, for halo masses $10^{10}$ (left), $10^{12}$ (middle), and
    $> 10^{14} M_\odot$ (right). Within each panel, the accretion rates at
    $z = 0, 0.5, 1$, and 2 are shown (except the right panel, where such
    massive haloes are present only at $z=0$ and 0.5). The distributions
    are seen to broaden significantly with increasing $z$. The vertical
    axis in each of the bottom panels labels the number of haloes per
    comoving Mpc$^3$ that are accreting at a rate of $\dot{M}_b$ or
    above.} \label{fig:MdotDist}
\end{figure*}

To compute the total mass growth rate of a halo of a given mass $M_0$ at
time $t$, we follow the main branch of its merger tree (see
\S~\ref{sec:extract}) and set $\dot{M}=(M_0-M_1)/\Delta t$, where $M_0$ is
the descendant mass at time $t$ and $M_1$ is the mass of its most massive
progenitor at time $t-\Delta t$. The mean value of $\dot{M}$ as a function
of $z$ for the complete set of resolved haloes in the two Millennium
simulations is plotted in Fig.~\ref{fig:Mdot} (solid curves). Nine ranges
of $M_0$ spanning five orders of magnitude ($10^{10} \,\msun$ to $10^{15}
\,\msun$ from bottom up) are shown. Fig.~\ref{fig:Mdot} can be compared
directly to Fig.~5 of \cite{MFM09} for the Millennium Simulation alone. The
rising $\mdot$ with increasing redshift in our earlier study is seen to
continue to $z\sim 14$, and the nearly linear scaling of $\mdot$ with halo
mass is extended down to $\sim 10^{10} M_\odot$.

We find the mass accretion rates shown in Fig.~\ref{fig:Mdot} to be very well fit by the forms given by equations~(8) and (9) of \citet{MFM09}. The coefficients quoted there only need minor adjustments after the Millennium-II results are added. We suggest the following updated fits to the mean and median mass growth rates of haloes of mass $M$ at redshift $z$:
\begin{eqnarray}
	\mdot_{\rm mean} &=& 46.1 \, \,\msun {\rm yr}^{-1} \left( \frac{M}{10^{12} \,\msun} \right)^{1.1} \nonumber\\
	&& \times (1 + 1.11 z) \sqrt{\Omega_m (1+z)^3 + \Omega_\Lambda} \nonumber\\
\label{Mdotfit} 
	\mdot_{\rm median} &=& 25.3 \, \,\msun {\rm yr}^{-1} \left( \frac{M}{10^{12} \,\msun} \right)^{1.1} \nonumber\\
	&& \times (1 + 1.65 z) \sqrt{\Omega_m (1+z)^3 + \Omega_\Lambda} \,. 
\end{eqnarray}
At a given mass and redshift, the mean rate is overall higher than the
median rate since the distribution of $\dot{M}$ has a long positive tail
(see Fig.~\ref{fig:MdotDist}). The dashed curves in Fig.~\ref{fig:Mdot}
illustrate the remarkable accuracy of this formula in matching the
simulation results over the broad ranges of halo mass and redshift shown.

The right-hand-side label along the vertical axis of Fig.~\ref{fig:Mdot}
shows the corresponding mean accretion rate of baryons, $\dot{M}_b$, where
we have assumed a cosmological baryon-to-dark matter ratio of
$\Omega_b/\Omega_m = 1/6$. These values are meant to provide a rough
approximation for the mean rate at which baryons are being accreted near
the virial radius of a dark matter halo.  Most of these baryons are
presumably in the form of warm or hot ionized hydrogen gas that is being
channeled into the haloes along cosmic filaments, and various gas
  cooling and feedback processes will likely affect the baryon accretion
  rate.  Many studies on galaxy formation are aimed at quantifying these
physical processes under which these baryons are cooled to form neutral
gas, molecular gas, and stars, and the feedback processes that heat up the
baryons and lead to large-scale outflows.

In Fig.~\ref{fig:MdotDist} we show the differential (top) and cumulative
(bottom) {\it distributions} of the baryonic accretion rate for three halo
masses (left to right panels) and four redshifts. A cosmic ratio of
$\Omega_b/\Omega_m=1/6$ is again assumed to convert the dark matter rate
into a baryonic rate.  The distributions are strongly peaked at the mean
values presented in Fig.~\ref{fig:Mdot} but exhibit long tails towards high
positive values due to major merger events and towards negative values due
to tidal stripping and halo fragmentation. Not only is the mean accretion
rate higher at higher $z$, the distribution of $\dot{M}_b$ is also broader
at higher $z$.  For example, the comoving density of Milky Way-mass haloes
that are accreting baryons at a rate of at least $100 M_\odot$ per year is
approximately ten times greater at $z=2$ ($\sim 2\times 10^{-4}$
Mpc$^{-3}$) than at $z=0$ ($\sim 3\times 10^{-5}$ Mpc$^{-3}$).

\subsubsection{Mass Assembly Histories}

\begin{figure}
  \centering
  \includegraphics{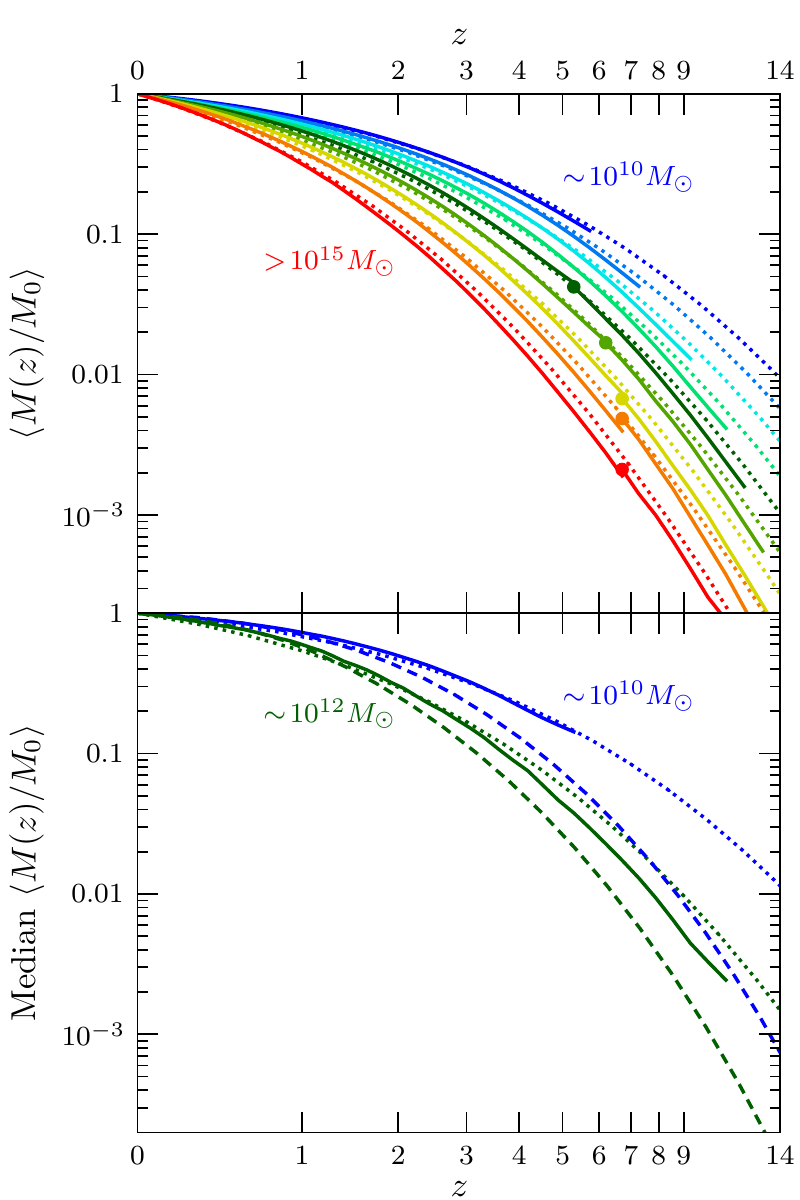} \caption{Top panel: {\it Mean} mass
    assembly history $M(z)$ of all $z=0$ resolved dark matter haloes in the
    two Millennium simulations (solid curves). Nine ranges of halo mass
    from $10^{10} M_\odot$ (top blue) to $10^{15} M_\odot$ (bottom red) are
    plotted. The dotted curves show the predictions given by integrating
    the mean $\dot{M}$ of our fitting formula (eq.~\ref{Mdotfit}). The
    lower four mass bins contain only haloes from the Millennium-II
    Simulation. For the upper five mass bins in which the haloes are drawn
    from both simulations, we use a solid circle to label the redshift
    above which only Millennium-II haloes contribute since the Millennium
    Simulation can no longer resolve haloes at such high $z$. The
    relatively smooth connection at the circle illustrates the consistency
    between the two simulations. Lower panel: {\it Median} mass assembly
    history $M(z)$.  For clarity, only two mass bins are plotted.  Solid
    lines are from the simulations, dotted lines from the integration of
    the \emph{mean} $\dot{M}$ from eq.~(\ref{Mdotfit}), and dashed lines
    show the fits from Zhao et al. (2009).}
    \label{fig:MZ}
\end{figure}

The top panel of Fig.~\ref{fig:MZ} shows the mean mass assembly history
$M(z)$ for nine bins of $M_0$ (at $z=0$) from $10^{10}$ to $>10^{14}$ (from
top to bottom).  The solid curves show the results obtained from the main
branch (i.e. the most massive progenitor) along the merger tree for all the
$z=0$ haloes in the two Millennium simulations.  The dotted curves show the
$M(z)$ obtained from integrating the fitting formula for the mean $\dot{M}$
in equation~(2) from the present-day to some redshift $z$.  The agreement
is generally very good, in particular at $z\la 8$.  A perfect agreement is
not expected because the two quantities, $\dot{M}$ and $M(z)$, are not
determined from the same set of haloes in the simulations: the $\dot{M}$
statistics are obtained from all haloes of a given mass $M$ at a given $z$,
whereas the $M(z)$ curves show only the mean mass of the most massive
progenitors at redshift $z$ for the $z=0$ haloes of mass $M$, which are a
small subset of the haloes of the same mass that are present at $z$ in the
simulation boxes. In view of this difference, the agreement between the
solid and dotted curves in the top panel of Fig.~\ref{fig:MZ} is in fact
quite remarkable. Over the large range of mass and redshift shown in
Fig.~\ref{fig:MZ}, we have checked that the direct fits for the mean $M(z)$
proposed in recent literature (e.g. \citealt{BK09b, MFM09}) provide a good
match at low $z$, but integrating the fit for $\mdot$ in
equation~(\ref{Mdotfit}) provides a closer match at high $z$,

The solid curves in the bottom panel of Fig.~\ref{fig:MZ} show the median,
rather than the mean, mass assembly history obtained from the simulations
for two mass bins centered at $M_0=10^{10} M_\odot$ and
$10^{12}M_\odot$.  We note that integrating our fit to the
\emph{median} $\dot{M}$ does not yield the median $M(z)$ because unlike the
\emph{mean} $\dot{M}$, the median and derivative operations do not commute.
The median and mean $M(z)$ are sufficiently similar, however, that we find
integrating our \emph{mean} $\dot{M}$ to yield relatively good agreement
with the median $M(z)$ (dotted curves).  For comparison, the fit of
\cite{Zhao09} to the median $M(z)$ is shown as dashed curves.  Their fit
appears to be systematically lower than the Millennium results at $z>1$.

 In principle, we can integrate the (mass-weighted) halo merger rate in
  equation~(1) and obtain the portion of the dark matter accretion rate
  $\mdot$ in equation~(2) that is due to mergers.  As emphasized in
  \cite{FM10}, however, accretion of ``diffuse'' material (consisting of
  unresolved haloes and tidally stripped mass) also makes a non-negligible
  contribution to $\mdot$; equation~(2) therefore can not be obtained
  solely from equation~(1).

\begin{figure*}
\includegraphics{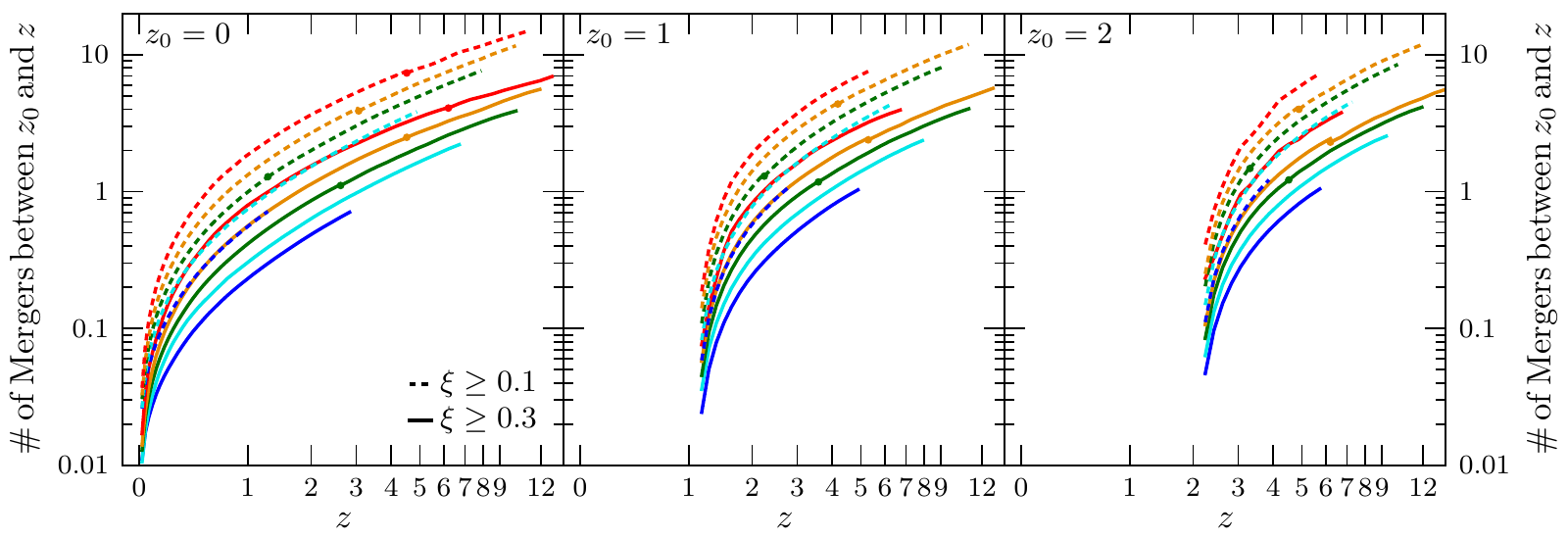} \caption{Mean number of mergers between
  redshifts $z_0$ and $z$ experienced by a halo at $z_0=0$ (left), 1
  (middle), and 2 (right) from the joint dataset of the two Millennium
  simulations. In each panel, the solid and dashed curves represent mergers
  with a progenitor mass ratio (defined at the time of merger) of $\xi \ge
  0.3$ and $\xi \ge 0.1$. For each mass ratio cutoff, five ranges of halo
  mass are shown (from bottom up): $10^{10}$ (blue), $10^{11}$ (cyan),
  $10^{12}$ (green), $10^{13}$ (orange), and $>10^{14} M_\odot$ (red). The
  lower-mass haloes are from the Millennium-II Simulation, whereas the
  cluster mass haloes are mainly from the Millennium Simulation. For mass
  bins in which the haloes are drawn from both simulations, we use a solid
  circle to label the redshift above which only Millennium-II haloes
  contribute since the Millennium can no longer resolve haloes at such high
  $z$. The fact that the curves connect quite smoothly are another
  indication of the consistency between the two
  simulations.} \label{fig:Nm}
\end{figure*}

\subsection{Merger Statistics over a Halo's History} \label{sec:mergerHistory}

In the last two sections we quantified the halo merger rates, the mass
growth rates, and the assembly histories of haloes. These quantities can be
combined to predict a number of additional useful merger statistics over a
halo's history.

\subsubsection{Cumulative number of mergers}

One such statistic is $N_m(\ximin, M_0,z_0,z)$, the total number of mergers
that a halo of mass $M_0$ at redshift $z_0$ has encountered between $z_0$
and an earlier $z$. The mergers can be characterized by major or minor mergers by
imposing a limit of $\ximin$ on the progenitor mass ratio (evaluated at the
redshift of the merger). These numbers are essential for making theoretical
predictions of galaxy properties that are impacted by mergers, e.g., the
dynamics and stability of stellar disks, the star formation rate, and the
color and morphology transformation due to mergers.

Fig.~\ref{fig:Nm} shows $N_m(\ximin, M_0,z_0,z)$ for the complete set of
resolved haloes at $z_0=0$ (left), 1 (middle), and 2 (right) from the two
Millennium simulations. In each panel, five ranges of $M_0$ are plotted for
redshift up to 12. Major mergers with $\xi \ge 0.3$ are shown by solid
curves, while the more minor mergers with $\xi \ge 0.1$ are shown in dashed
curves.

Fig.~\ref{fig:Nm} shows that the mean trend of the number of mergers
experienced over a halo's lifetime is a sensitive function of the halo mass
and merger mass ratio. haloes of Milky-Way mass at the present day (green
curves) have on average experienced one major merger event ($\xi \ge 0.3$)
per halo since $z\approx 2.3$, and one merger with $\xi \ge 0.1$ per halo
since $z\approx 1$. When extended to $z\approx 7$, these haloes have on
average encountered $\sim 3$ mergers with $\xi \ge 0.3$, and $\sim 7$
mergers with $\xi \ge 0.1$. The formation redshifts as well as the last
merger epoch for more massive haloes are both lower, a well-known
  trend in CDM-based cosmology (see, e.g., \citealt{LC93,LC94}).
Cluster-sized haloes with $M_0\sim 10^{14} M_\odot$, for instance, have on
average experienced one major merger event since $z\approx 1.2$, and one
merger with $\xi \ge 0.1$ since $z\approx 0.6$.

\begin{figure}
  \includegraphics{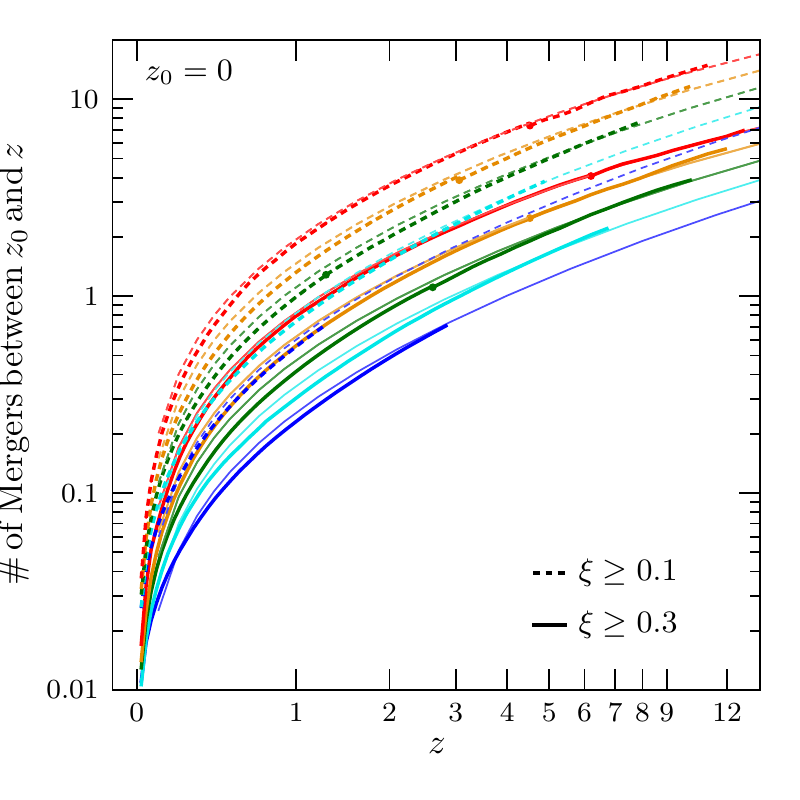} \caption{Same as the left panel of
    Fig.~\ref{fig:Nm}, with the addition of the predictions (thin curves)
    computed from eq.~(\ref{Nm}).  The agreement with the simulation
    results (thick curves) is excellent, suggesting that eq.~(\ref{Nm}) can
    be used to make analytic predictions for merger statistics over a
    halo's history.  }
\label{fig:Nmz0}
\end{figure}

It is possible to compute the cumulative number of mergers,
$N_m(\ximin,M_0,z_0,z)$, shown in Fig.~\ref{fig:Nm} from the
fitting formula for the merger rate $dN_m/d\xi/dz$ in equation~(\ref{fit})
and the mass accretion history $M(z)$ obtained by integrating $\mdot$ in
equation~(\ref{Mdotfit}).  Specifically, these quantities are related by
\begin{equation}
	N_m(\ximin, M_0,z_0,z) = \int_{z_0}^z dz \int_{\ximin}^1 d\xi 
    \frac{dN_m}{d\xi dz}\left( M(z),\xi,z \right) \,. 
\label{Nm} 
\end{equation}
Since we are interested in the number of mergers over a halo's past
history, we must take into account the fact that a halo's mass generally
decreases with increasing $z$, and that the merger rate depends on the halo
mass (albeit weakly). The merger rate $dN_m/d\xi/dz$ at redshift $z$ in the
integrand above therefore should be evaluated using the mean mass $M(z)$
that a halo of mass $M_0$ at $z_0$ had at the earlier $z$.  The results are
shown in Fig.~\ref{fig:Nmz0}, which is identical to the left panel of
Fig.~\ref{fig:Nm} except that we have added the theoretical curves (thin
curves) for comparison.  The agreement with the simulation results (thick
curves) is excellent, suggesting that equation~(\ref{Nm}) can be used to
make analytic predictions for merger statistics over a halo's history.

\begin{figure*}
  \includegraphics{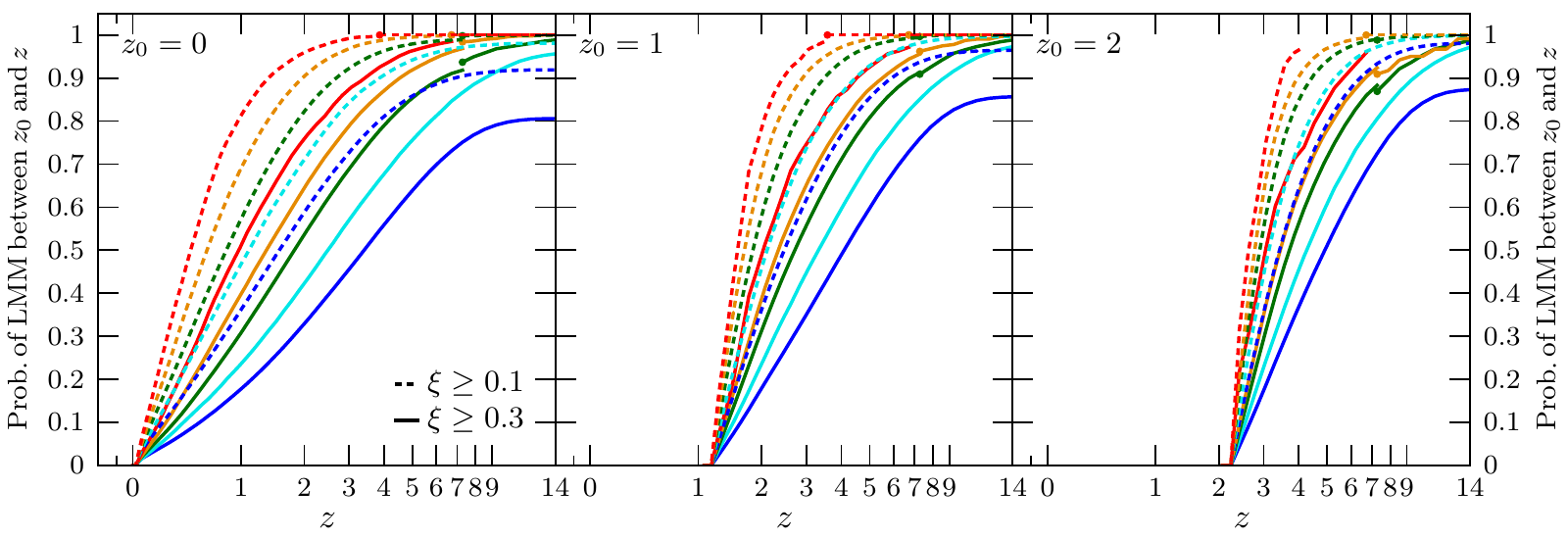} \caption{Cumulative distribution of
    the redshift at which the last (i.e. most recent) major merger occurred
    in a halo's past history for haloes at $z_0=0$, 1, and 2 (left to
    right) in the two Millennium simulations. The vertical axis gives the
    probability that the last major merger occurred between $z_0$ and
    redshift $z$. The curves are defined the same way as in
    Fig.~\ref{fig:Nm}. } \label{fig:LMM}
\end{figure*}

\subsubsection{Redshift of last major merger}

The redshift at which each curve in Fig.~\ref{fig:Nm} crosses one merger
event along the vertical axis is a useful quantity since it gives the mean
redshift at which a halo has experienced its last major merger (LMM). The
LMM redshift of a halo is closely related to its formation redshift and
may be linked to the time at which the associated galaxy last
experienced prominent star formation activity and morphological changes. To
analyze this quantity further, we show in Fig.~\ref{fig:LMM} the {\it
  distribution} of the LMM redshift for haloes at $z_0=0$, 1, and 2 (from
left to right). Within each panel, five halo masses and two ranges of $\xi$
are plotted. The vertical axis gives the probability that a halo at a given
$z_0$ has had a last major merger between $z_0$ and $z$.

Useful merger statistics can be read off from Fig.~\ref{fig:LMM}. For
instance, 50\% of present-day haloes have had a major merger ($\xi \ge
0.3$) since $z\approx 1$, 1.8, and 3.4 for halo mass $10^{14}, 10^{12}$,
and $10^{10} M_\odot$, respectively. When more minor mergers with $\xi \ge
0.1$ are considered, the median redshift of the last merger is
lowered to 0.4, 0.8, and 1.6 for the three masses. The assembly history of
Milky Way-size haloes is of particular interest; see \cite{BK09b} and
references therein for a detailed statistical study of this topic. For
haloes of $M_0 \sim 10^{12} M_\odot$ today, the left panel of
Fig.~\ref{fig:LMM} shows that $\sim 31\%$, 53\%, and 69\% of these haloes
have experienced a major merger since $z=1$, 2, and 3, respectively. For
haloes of $\sim 10^{12} M_\odot$ at $z_0=1$ (middle panel), about 50\% of
them have had a major merger since $z\approx 2.7$, and for haloes of the
same mass at $z_0=2$ (right panel), about 50\% of them have had a major
merger since $z\approx 3.7$.

\section{Summary and Conclusions} \label{sec:conclusions}

We have combined the halo catalogs from the two Millennium simulations to
form an unprecedentedly large dataset for studying the merger statistics
and assembly histories of dark matter haloes in the $\Lambda$CDM
cosmology. The two simulations provide, respectively, $11.3\times 10^{6}$
haloes (between redshift 0 and 6) and $7.5\times 10^{6}$ haloes (between
redshift 0 and 15) above 1000 particles for our study. These haloes and
their merger trees have allowed us to determine the dark matter halo merger
rates and mass growth rates from $z=0$ to up to $z=15$, for over five
orders of magnitude in the descendant halo mass ($10^{10} \la M_0 \la
10^{15} M_\odot$) and progenitor mass ratio ($10^{-5} \la \xi \le 1$). For
the small range of overlapping parameter space between the two simulations,
we have found the agreement to be excellent.

For the merger rates, the basic features reported in our earlier study based on the Millennium Simulation alone \citep{FM08} are largely preserved in the Millennium-II Simulation. The mean merger rate {\it per halo}, $dN_m/d\xi/dz$, is nearly independent of the descendant mass (Fig.~\ref{fig:Collapsing} and \ref{fig:MassDependence}) and scales as $\propto M_0^{0.133}$ at all redshifts. The merger rate in units of per redshift is nearly independent of redshift out to $z\sim 15$ (left panel of Fig.~\ref{fig:BNZ}); the rate in units of per Gyr is therefore largely determined by the cosmological factor of $dt/dz$ and increases roughly as $(1+z)^{2.5}$ at $z\ga 1$ (right panel of Fig.~\ref{fig:BNZ}). Equation~(\ref{fit}) provides an update on our simple analytical fitting form for the merger rate as a function of $M_0$, $\xi$, and $z$.

For the mass growth rates of individual haloes, we have found the mean and median statistics (Fig.~\ref{fig:Mdot}) to be well approximated by the simple fitting form of our earlier study \citep{MFM09}. The updated coefficients based on the joint dataset from the two Millennium simulations are given by equation~(\ref{Mdotfit}). The present-day mean and median rates at which a $10^{12} M_\odot$ dark matter halo is accreting mass (at the virial radii) are 46.1 and 25.3 $M_\odot$ yr$^{-1}$, respectively. This rate increases nearly linearly with the halo mass, and increases with redshift approximately as $(1+z)^{1.5}$ at low $z$ and $(1+z)^{2.5}$ at $z\ga 1$.

We have also presented statistical quantities that track the merger histories of dark matter haloes cumulatively. Fig.~\ref{fig:Nm} presents the number of major mergers experienced by haloes of various mass between redshift $z_0$ and $z$ for $z_0=0,1,$ and 2. Fig. ~\ref{fig:LMM} presents the probability that a dark matter halo at redshift $z_0$ will have last experienced a major merger at some earlier redshift $z$. Much interesting and useful information regarding the contribution to halo growth made by major mergers can be read off these figures with ease.

With the addition of results from the Millennium-II Simulation to our previous analysis of the Millennium Simulation, the merger rate of dark matter haloes is now well-quantified for haloes with masses between $10^{10}$ and $10^{15}\,\msun$ for redshifts $z \la 15$, modulo the uncertainties inherent in halo definitions and in algorithms for handling fragmentation (see Appendix), for the cosmology used in the Millennium simulations. Several avenues remain open for future work, however.

One obvious extension of the results in this paper is to consider the mergers of {\it subhaloes} themselves, as subhalo mergers can be more directly linked to galaxy mergers than can FOF halo mergers \citep{Angulo09,Wetzel09}. Furthermore, the structure of the merger trees produced for the Millennium simulations lends itself naturally to computing subhalo merger properties. While computing subhalo merger rates and connecting them to galaxy mergers presents additional challenges -- in particular, the issues of assigning stellar masses to subhaloes, numerical resolution effects, and subhalo identification within larger FOF haloes -- a thorough theoretical understanding of such rates is essential for disentangling the relative contributions of merging and star formation to the growth of galaxies.

\section*{Acknowledgements}

The Millennium Simulation databases used in this paper and the web application providing online access to them were constructed as part of the activities of the German Astrophysical Virtual Observatory.

\section*{Appendix: Comparison of Different FOF Merger Trees} \label{sec:appendix}

We refer the reader to Section~5 and Figure~8 of \cite{FM09} for a
detailed discussion of the three basic operations -- ``snip,'' ``stitch,''
and ``snip'' -- that we have implemented and tested for handling the issue
of halo fragmentations during the construction of a merger tree for FOF
haloes (see also Section~2.2 of this paper).  Briefly, ``snip''
  removes halo fragmentation events by severing the ancestral link between
  the fragment subhalo and its progenitor, ``stitch'' places the fragment
  subhalo back into the FOF halo from which it emerged, whereas ``split''
  removes the fragment subhalo's progenitor from its FOF halo, thereby
  generating a new FOF at the progenitor redshift.

Within the stitch and split algorithms, the operations can be applied
either on a subset of fragments or on all fragments in a given FOF tree. We
therefore subdivide each algorithm into two: stitch-$\infty$ vs stitch-3,
and split-$\infty$ vs split-3. The stitch-$\infty$ and split-$\infty$
algorithms perform the given operation on all FOF fragments. This is done
recursively from the redshift of fragmentation, going forward in redshift
for stitch-$\infty$, and backwards in redshift for split-$\infty$, until
there are no more fragments present in the simulation merger trees. As a
result of this recursive process, stitch-$\infty$ identifies the first
(highest-z) snapshot in which two subhaloes join the same FOF to be their
merger time, whereas split-$\infty$ selects the last (lowest-z) snapshot.

These algorithms introduce some complications, however. One particular problem faced by split-$\infty$ is the fact that there exists a firm cutoff at $z=0$, beyond which we do not have merger or fragmentation information. As a result, although a fragment may actually finally merge beyond $z=0$, split-$\infty$ will incorrectly assign its final merger to an earlier redshift. This results in a pile-up of merger events at $z=0$ and, as we will show, artificially raises the low-$z$ merger rate with respect to the high-$z$ rate. Since there is no analogous hard limit at high $z$, stitch-$\infty$ does not suffer from this same behavior, and fragment mergers are re-distributed across all high redshifts evenly. On the other hand, any chance encounter between subhaloes that results in the FOF algorithm spuriously linking them together is interpreted as a real merger event by stitch-$\infty$. The subhaloes, which may never interact again, are nonetheless forced to join the same FOF group down to $z=0$.

The stitch-3 and split-3 algorithms are designed to limit the propagation effects of stitch-$\infty$ and split-$\infty$. Stitch-3 performs the stitching operation on any FOF fragment that is observed to remerge with its progenitor FOF's main branch within 3 snapshots of the fragmentation event. Any fragments that do not satisfy this criterion are snipped, resulting in an orphan halo that may or may not later remerge. The split-3 algorithm performs the split operation on any FOF fragment that is \emph{not} a member of the main branch FOF at some point in the 3 snapshots \emph{before} the fragmentation event. Again, fragments that do not satisfy this criterion are snipped.

Neither stitch-3 nor split-3 adequately removes all remerger events. Depending on the context this may be either a weakness or a strength: although the notion of halo remergers may be considered as multiple counting from a theoretical perspective, observers will likely count as signatures all events that trigger mergers, regardless of whether they are the first or last entry.

Moreover, both split-3 and stitch-3 have superior time convergence properties to the snip algorithm, in which the remerger problem is entirely unmitigated. Thus, stitch-3 and split-3 stand as intermediates between the snip and stitch-$\infty$/split-$\infty$ algorithms. 
\begin{figure*}
	\centering 
	\includegraphics{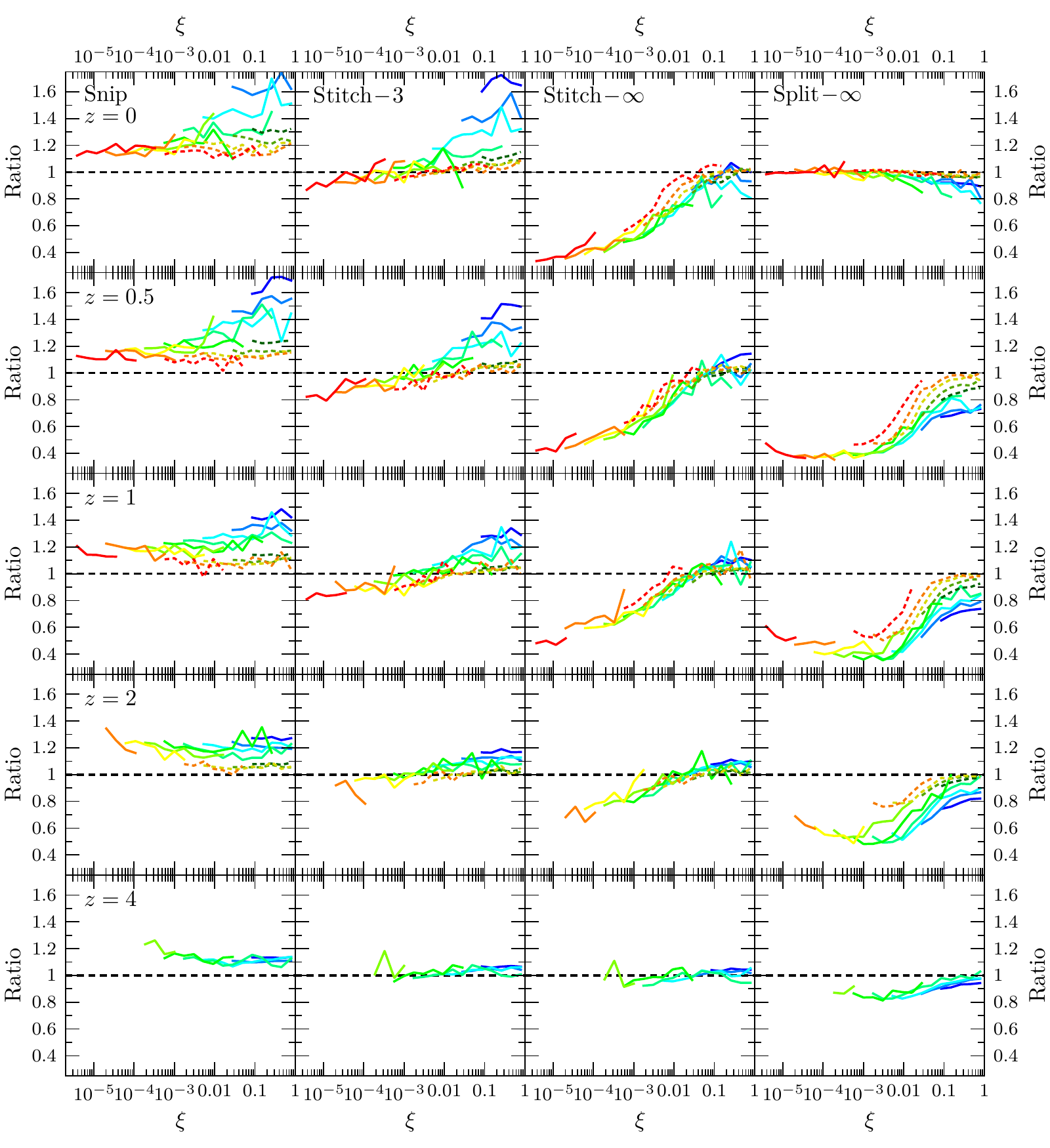} \caption{Comparison of five
          algorithms used to handle halo fragmentation events in the FOF
          merger trees: snip, stitch-3, stitch-$\infty$, split-$\infty$,
          and split-3. Results presented throughout this paper are based on
          the split-3 tree. Plotted as a function of the progenitor mass
          ratio $\xi$ (left to right) are the ratios of the merger rates,
          $dN_m/d\xi/dz$, between each of the first four algorithms
          relative to the split-3 results. Five redshifts are shown:
          $z=0,0.5,1,2$ and $4$ (top to bottom). Within each panel, up to
          nine mass bins are shown: $10^{10} M_\odot$ (blue) to $>10^{14}
          M_\odot$ (red). The Millennium Simulation results are presented
          with dashed curves and Millennium II with solid curves. The
          systematic differences amongst the five algorithms are discussed
          in the text.} \label{fig:BNRatio}
\end{figure*}

An immediate concern is whether the halo mass function is heavily modified by the destruction/creation of FOFs due to the stitch/split operations. We have verified that these operations do not modify the mass function severely. For stitch-3 and split-3, the deviations are within 3\% of the unprocessed mass function at all redshifts, while deviations of up to 10\% exist for the stitch-$\infty$ (split-$\infty$) algorithm at low (high) redshifts.

Fig.~\ref{fig:BNRatio} compares the five post-processing algorithms
directly by presenting ratios of the per-halo merger rate, $dN_m/d\xi/dz$, as a
function of progenitor mass ratio $\xi$ at five redshifts ($z=0.06$, 0.5,
1, 2, and 4 from top bottom). Each column presents the ratio of the merger
rate of a particular post-processing algorithm (left to right: snip,
stitch-3, stitch-$\infty$, split-$\infty$) to the merger rate extracted
from the split-3 trees presented throughout this paper. Different mass bins
are presented by different colored curves ranging from $10^{10} \,\msun$
(blue) to $10^{15} \,\msun$ (red). We note that though the region of
overlap between Millennium (dashed) and Millennium II (solid) is small,
there appears to be smooth continuation between these two sets of curves
for all post-processing algorithms.

The merger rates computed by all algorithms converge towards high $z$, though there is some residual disagreement with split-$\infty$ and snip at the $\sim 20\%$ level. There are, however, distinct systematic differences among the algorithms when $z<4$. Since we presented stitch-3 as our algorithm of choice for handling halo fragmentation in \cite{FM08}, we focus on the comparison of stitch-3 and split-3 in this section. The origins of the differences between the other algorithms and split-3 can be inferred from the discussion of the algorithms earlier in this section.

The second column of Fig.~\ref{fig:BNRatio} shows that stitch-3 and split-3 are in excellent agreement at all $\xi$ for high mass haloes ($M_0>10^{12}$). Low mass haloes, however, can show significant deviations in the merger rate. This is true especially in the major merger regime, where the merger rate predicted by stitch-3 is over $50\%$ higher than split-3. This distinction was not detectable using Millennium alone, as the mass resolution limited our analysis to $M_0 \geq 10^{12} M_\odot$.

To understand this deviation we have studied a subset of halo mergers in detail by analyzing halo tracks, velocities, and merger histories. In particular, we have constructed a number of criteria to determine whether a given merger is actually a spurious encounter: if the relative velocity of the two haloes greatly exceeds the more massive halo's maximum circular velocity, if the angle between the velocity vectors of the two haloes exceeds $70^{\circ}$ at the time of merger, or if the FOF algorithm only associates the two haloes for two snapshots out of the eight snapshots centered on the merger snapshot, then the merger is deemed spurious. A qualitative look at three-dimensional halo trajectories finds that this criteria does a good job of identifying chance halo encounters and premature mergers.

For halo mergers with $1.1\times10^{10}<M_0<1.3\times 10^{10}$ and $\xi>0.1$ at $z=0$, we find that stitch-3 identifies $1,304$ mergers, while split-3 only identifies $738$ mergers. Of these, 505 mergers are in common, leaving split-3 with 233 mergers that are not in stitch-3 and stitch-3 with 799 mergers that are not in split-3. Of the 505 mergers in common, only 4 (0.8\%) are deemed spurious by our criterion. Similarly, of the 233 mergers unique to split-3, only 12 (5.2\%) are deemed spurious. Of the 799 mergers unique to stitch-3, however, 589 (73.7\%) are deemed spurious. These spurious mergers are primarily comprised of chance encounters in which the two otherwise unassociated haloes merge for a snapshot or two and then disconnect. While split-3 correctly splits these events, stitch-3 does not and consequently inflates the merger rate. When these spurious mergers are removed, the remaining 210 mergers unique to the stitch-3 algorithm bring the stitch-3 and split-3 rates into close agreement.

We note that depending on the context, one may choose one algorithm over another. Stitch-$\infty$ provides the \emph{first} encounter merger rate, but is known to link chance-encounter haloes that should not be linked. Split-$\infty$ provides the \emph{last} encounter merger rate, but cannot be trusted for $z<1$ and may incorrectly underpredict the merger rate due to spurious fragmentation. Split-3 stands in between both algorithms: it does not propagate up and down the tree and does not heavily modify the distribution of FOFs, but it does double count some halo remerger events. This may be odious to the theorist, but may yield the most appropriate merger rate for comparison to observation.

\bibliographystyle{mn2e} 
\bibliography{FMBK10}

\begin{thebibliography}{}

\bibitem[\protect\citeauthoryear{{Angulo}, {Lacey}, {Baugh} \&
  {Frenk}}{{Angulo} et~al.}{2009}]{Angulo09}
{Angulo} R.~E.,  {Lacey} C.~G.,  {Baugh} C.~M.,    {Frenk} C.~S.,  2009,
  \mnras, 399, 983

\bibitem[\protect\citeauthoryear{{Berrier}, {Bullock}, {Barton}, {Guenther},
  {Zentner} \& {Wechsler}}{{Berrier} et~al.}{2006}]{Berrier06}
{Berrier} J.~C.,  {Bullock} J.~S.,  {Barton} E.~J.,  {Guenther} H.~D.,
  {Zentner} A.~R.,    {Wechsler} R.~H.,  2006, \apj, 652, 56

\bibitem[\protect\citeauthoryear{{Boylan-Kolchin}, {Springel}, {White} \&
  {Jenkins}}{{Boylan-Kolchin} et~al.}{2009}]{BK09b}
{Boylan-Kolchin} M.,  {Springel} V.,  {White} S.~D.~M.,    {Jenkins} A.,  2009,
  arXiv:0911.4484 [astro-ph]

\bibitem[\protect\citeauthoryear{{Boylan-Kolchin}, {Springel}, {White},
  {Jenkins} \& {Lemson}}{{Boylan-Kolchin} et~al.}{2009}]{BK09}
{Boylan-Kolchin} M.,  {Springel} V.,  {White} S.~D.~M.,  {Jenkins} A.,
  {Lemson} G.,  2009, \mnras, 398, 1150

\bibitem[\protect\citeauthoryear{{Cole}, {Helly}, {Frenk} \&
  {Parkinson}}{{Cole} et~al.}{2008}]{Cole08}
{Cole} S.,  {Helly} J.,  {Frenk} C.~S.,    {Parkinson} H.,  2008, \mnras, 383,
  546

\bibitem[\protect\citeauthoryear{{Davis}, {Efstathiou}, {Frenk} \&
  {White}}{{Davis} et~al.}{1985}]{Davis85}
{Davis} M.,  {Efstathiou} G.,  {Frenk} C.~S.,    {White} S.~D.~M.,  1985, ApJ,
  292, 371

\bibitem[\protect\citeauthoryear{{Fakhouri} \& {Ma}}{{Fakhouri} \&
  {Ma}}{2010}]{FM10}
{Fakhouri} O.,  {Ma} C.,  2010, \mnras, 401, 2245

\bibitem[\protect\citeauthoryear{{Fakhouri} \& {Ma}}{{Fakhouri} \&
  {Ma}}{2008}]{FM08}
{Fakhouri} O.,  {Ma} C.-P.,  2008, \mnras, 386, 577

\bibitem[\protect\citeauthoryear{{Fakhouri} \& {Ma}}{{Fakhouri} \&
  {Ma}}{2009}]{FM09}
{Fakhouri} O.,  {Ma} C.-P.,  2009, \mnras, 394, 1825

\bibitem[\protect\citeauthoryear{{Genel}, {Genzel}, {Bouch{\'e}}, {Naab} \&
  {Sternberg}}{{Genel} et~al.}{2009}]{Genel09}
{Genel} S.,  {Genzel} R.,  {Bouch{\'e}} N.,  {Naab} T.,    {Sternberg} A.,
  2009, \apj, 701, 2002

\bibitem[\protect\citeauthoryear{{Gottl{\"o}ber}, {Klypin} \&
  {Kravtsov}}{{Gottl{\"o}ber} et~al.}{2001}]{Gottlober01}
{Gottl{\"o}ber} S.,  {Klypin} A.,    {Kravtsov} A.~V.,  2001, ApJ, 546, 223

\bibitem[\protect\citeauthoryear{{Governato}, {Gardner}, {Stadel}, {Quinn} \&
  {Lake}}{{Governato} et~al.}{1999}]{Governato99}
{Governato} F.,  {Gardner} J.~P.,  {Stadel} J.,  {Quinn} T.,    {Lake} G.,
  1999, \aj, 117, 1651

\bibitem[\protect\citeauthoryear{{Guo} \& {White}}{{Guo} \&
  {White}}{2008}]{Guo08}
{Guo} Q.,  {White} S.~D.~M.,  2008, \mnras, 384, 2

\bibitem[\protect\citeauthoryear{{Lacey} \& {Cole}}{{Lacey} \&
  {Cole}}{1993}]{LC93}
{Lacey} C.,  {Cole} S.,  1993, MNRAS, 262, 627

\bibitem[\protect\citeauthoryear{{Lacey} \& {Cole}}{{Lacey} \&
  {Cole}}{1994}]{LC94}
{Lacey} C.,  {Cole} S.,  1994, MNRAS, 271, 676

\bibitem[\protect\citeauthoryear{{Li}, {Mo}, {van den Bosch} \& {Lin}}{{Li}
  et~al.}{2007}]{Li07}
{Li} Y.,  {Mo} H.~J.,  {van den Bosch} F.~C.,    {Lin} W.~P.,  2007, \mnras,
  379, 689

\bibitem[\protect\citeauthoryear{{Maller}, {Katz}, {Kere{\v s}}, {Dav{\'e}} \&
  {Weinberg}}{{Maller} et~al.}{2006}]{Maller06}
{Maller} A.~H.,  {Katz} N.,  {Kere{\v s}} D.,  {Dav{\'e}} R.,    {Weinberg}
  D.~H.,  2006, \apj, 647, 763

\bibitem[\protect\citeauthoryear{{McBride}, {Fakhouri} \& {Ma}}{{McBride}
  et~al.}{2009}]{MFM09}
{McBride} J.,  {Fakhouri} O.,    {Ma} C.,  2009, \mnras, 398, 1858

\bibitem[\protect\citeauthoryear{{Springel}}{{Springel}}{2005}]{Springel05gadg%
et}
{Springel} V.,  2005, \mnras, 364, 1105

\bibitem[\protect\citeauthoryear{{Springel}, {White}, {Jenkins}, {Frenk},
  {Yoshida}, {Gao}, {Navarro}, {Thacker}, {Croton}, {Helly}, {Peacock}, {Cole},
  {Thomas}, {Couchman}, {Evrard}, {Colberg} \& {Pearce}}{{Springel}
  et~al.}{2005}]{Springel05}
{Springel} V.,  {White} S.~D.~M.,  {Jenkins} A.,  {Frenk} C.~S.,  {Yoshida} N.,
   {Gao} L.,  {Navarro} J.,  {Thacker} R.,  {Croton} D.,  {Helly} J.,
  {Peacock} J.~A.,  {Cole} S.,  {Thomas} P.,  {Couchman} H.,  {Evrard} A.,
  {Colberg} J.,    {Pearce} F.,  2005, Nat, 435, 629

\bibitem[\protect\citeauthoryear{{Springel}, {White}, {Tormen} \&
  {Kauffmann}}{{Springel} et~al.}{2001}]{springel01SUBFIND}
{Springel} V.,  {White} S.~D.~M.,  {Tormen} G.,    {Kauffmann} G.,  2001,
  MNRAS, 328, 726

\bibitem[\protect\citeauthoryear{{Springel}, {Yoshida} \& {White}}{{Springel}
  et~al.}{2001}]{Springel01}
{Springel} V.,  {Yoshida} N.,    {White} S.~D.~M.,  2001, New Astronomy, 6, 79

\bibitem[\protect\citeauthoryear{{Stewart}, {Bullock}, {Barton} \&
  {Wechsler}}{{Stewart} et~al.}{2009}]{Stewart09}
{Stewart} K.~R.,  {Bullock} J.~S.,  {Barton} E.~J.,    {Wechsler} R.~H.,  2009,
  \apj, 702, 1005

\bibitem[\protect\citeauthoryear{{Stewart}, {Bullock}, {Wechsler}, {Maller} \&
  {Zentner}}{{Stewart} et~al.}{2008}]{Stewart08}
{Stewart} K.~R.,  {Bullock} J.~S.,  {Wechsler} R.~H.,  {Maller} A.~H.,
  {Zentner} A.~R.,  2008, \apj, 683, 597

\bibitem[\protect\citeauthoryear{{Tormen}}{{Tormen}}{1998}]{Tormen98}
{Tormen} G.,  1998, \mnras, 297, 648

\bibitem[\protect\citeauthoryear{{Tormen}, {Bouchet} \& {White}}{{Tormen}
  et~al.}{1997}]{Tormen97}
{Tormen} G.,  {Bouchet} F.~R.,    {White} S.~D.~M.,  1997, \mnras, 286, 865

\bibitem[\protect\citeauthoryear{{van den Bosch}}{{van den
  Bosch}}{2002}]{VDBosch}
{van den Bosch} F.~C.,  2002, MNRAS, 331, 98

\bibitem[\protect\citeauthoryear{{Wechsler}, {Bullock}, {Primack}, {Kravtsov}
  \& {Dekel}}{{Wechsler} et~al.}{2002}]{WechslerMz}
{Wechsler} R.~H.,  {Bullock} J.~S.,  {Primack} J.~R.,  {Kravtsov} A.~V.,
  {Dekel} A.,  2002, ApJ, 568, 52

\bibitem[\protect\citeauthoryear{{Wetzel}, {Cohn} \& {White}}{{Wetzel}
  et~al.}{2009}]{Wetzel09}
{Wetzel} A.~R.,  {Cohn} J.~D.,    {White} M.,  2009, \mnras, 395, 1376

\bibitem[\protect\citeauthoryear{{Zhao}, {Jing}, {Mo} \& {B{\"o}rner}}{{Zhao}
  et~al.}{2009}]{Zhao09}
{Zhao} D.~H.,  {Jing} Y.~P.,  {Mo} H.~J.,    {B{\"o}rner} G.,  2009, \apj, 707,
  354

\end{thebibliography}

\label{lastpage}

\end{document}